\documentclass{article}%
\usepackage{amsmath}
\usepackage{amsfonts}
\usepackage{amssymb}
\usepackage{graphicx}%
\newtheorem{theorem}{Theorem}
\newtheorem{acknowledgement}[theorem]{Acknowledgement}

\begin{document}

\title{{\LARGE Semiclassical and quantum description of motion on noncommutative
plane}}
\author{M.C. Baldiotti\thanks{E-mail: baldiott@fma.if.usp.br}, J.P.
Gazeau\thanks{Laboratoire APC, Universit\'{e} Paris Diderot \textit{Paris 7}
10, rue A. Domon et L. Duquet 75205 Paris Cedex 13, France, E-mail:
gazeau@apc.univ-paris7.fr}, and D.M. Gitman\thanks{E-mail:
gitman@dfn.if.usp.br}\\Instituto de F\'{\i}sica, Universidade de S\~{a}o Paulo,\\Caixa Postal 66318-CEP, 05315-970 S\~{a}o Paulo, S.P., Brazil}
\maketitle

\begin{abstract}
We study the canonical and the coherent state quantization of a particle
moving in a magnetic field on a non-commutative plane. Starting from the so
called $\theta$-modified action, we perform the canonical quantization and
analyze the gauge dependence of the obtained quantum theory. We construct the
Malkin-Man'ko coherent states of the system in question, and the corresponding
quantization. On this base, we study the relation between the coherent states
and the \textquotedblleft classical\textquotedblright\ trajectories predicted
by the $\theta$-modified action. In addition, we construct different
semiclassical states, making use of special properties of circular squeezed
states. With the help of these states, we perform the Berezin-Klauder-Toeplitz
quantization and present a numerical exploration of the semiclassical behavior
of physical quantities in these states.

\end{abstract}

\section{Introduction}

Constructing semiclassical states for a given quantum system is an important
and, in general case, an open problem in quantum theory. One can believe that
for systems with quadratic Hamiltonians semiclassical (or \textquotedblleft
classical-like") states are those ones introduced by Shr\"{o}dinger
\cite{Sch26} in 1926, lately rediscovered and called coherent by Glauber
\cite{Gla63} and Sudarshan \cite{Sud63} within the context of Quantum Optics,
and by Klauder \cite{Klau60,Klau63} in the more general quantum arena. We will
call them standard CS in what follows. These states were then studied by a
number of authors in different contexts \cite{ZhaFeG90,KlaSk85,AliAnG00}.
Perelomov proposed so-called generalized CS for systems with a symmetry group
\cite{Per72}. One can also mention some alternative constructions of CS that
differ from the standard ones, see e.g. \cite{AliAnG00}. In addition, it was
realized that constructing CS is closely related to the quantization problem
\cite{Ber2}.

A charged quantum particle interacting with a constant magnetic field is an
important and well studied system. A family of coherent states adapted to such
a system was first proposed by Malkin and Man'ko \cite{MalMa68}. Afterwards,
some alternative constructions were proposed in \cite{LoyMoS89,SchMo03}.
However, all these CS are labelled not only by continuous quantum numbers, but
also by a discrete quantum number. To avoid discrete quantum numbers, the
authors of \cite{KowRe05} have proposed new coherent states partially similar
to the CS of a particle moving on a circle \cite{KowReP96}. A generalization
of such circular CS was proposed in our work \cite{GazBaG09}. Recently,
quantum states of a particle in a magnetic field and on a noncommutative plane
have been attracting considerable attention, see for instance
\cite{Hor02,AlGoKP08,GelGaS09} and \cite{DelDUGL07}. Among the problems in the
formulation of the non-commutative quantum mechanic are to implement the
symmetries of the ordinary theory and the consistent interpretation of the
position operator \cite{SchGoHR08}. Some works are devoted to the search for
experimental observation that can gives a physical evidences of the
non-commutative properties of the space, or that can be used to set some
limits for this non-commutative properties \cite{Chaichian1,AdoBaCGT09}. The
non-commutative quantum mechanics can be constructed by the quantization of a
classical $\theta$-modified action \cite{GitKu08}. In this article we will
construct the standard CS for a particle in a magnetic field and on a
non-commutative plane, and study the relation between these classical-like
states and the \textquotedblleft classical\textquotedblright\ trajectories
predicted by the corresponding $\theta$-modified action. In addition, we make
use of the interesting properties of the circular CS proposed in
\cite{GazBaG09} to construct semiclassical states of this system.

The present article is organized as follows. In Section 2, we recall the
formulation of classical mechanics on a non-commutative space in terms of
commuting coordinates (and in terms of the so-called $\theta$-modified
actions). In Section 3, we study such a formulation and its quantum version
for a charged particle submitted to a magnetic field, and the related gauge
dependence of this formulation. In Section 4, we follow the Malkin-Man'ko
approach and construct coherent states for the above system on the
non-commutative plane. We discuss their semiclassical properties and the
corresponding quantization of physical quantities. In Section 5, we construct
for the same system partially circular CS, proposed by us in \cite{GazBaG09},
and perform the Berezin-Klauder-Toeplitz quantization based precisely on these
coherent states. In addition, we present a numerical exploration of their
semiclassical behavior. Finally, in Section 6, we summarize the obtained results.

\section{Classical and Quantum motion on noncommutative plane}

Let us start by briefly recalling the nonrelativistic quantum mechanical
description of a finite-dimensional system living on a noncommutative space.
Suppose that such a system is described by coordinates $\hat{q}^{k}$ and
momenta $\hat{p}_{j}$ operators, $k,j=1,...,d,$ that obey the commutation
relations
\begin{equation}
\left[  \hat{q}^{k},\hat{q}^{j}\right]  =i\theta^{kj}\,,\;\left[  \hat{q}%
^{k},\hat{p}_{j}\right]  =i\hbar\delta_{j}^{k}\,,\;\left[  \hat{p}_{k},\hat
{p}_{j}\right]  =0\,,\;\theta^{kj}=-\theta^{jk}~, \label{1}%
\end{equation}
where $\theta^{kj}$ is a real constant antisymmetric matrix. The quantum
Hamiltonian $\hat{H}=\left.  H\left(  p,q\right)  \right\vert _{p\rightarrow
\hat{p},q\rightarrow\hat{q}}$ is constructed from the classical one $H\left(
p,q\right)  $ and a certain ordering is chosen. Introducing new operators, see
\cite{Chaichian1},%
\begin{align}
&  \hat{x}^{k}=\hat{q}^{k}+\frac{1}{2\hbar}\theta^{kj}\hat{p}_{j}~,\nonumber\\
&  \left[  \hat{x}^{k},\hat{x}^{j}\right]  =0\ ,\;\left[  \hat{p}_{k},\hat
{p}_{j}\right]  =0\ ,\;\left[  \hat{x}^{k},\hat{p}_{j}\right]  =i\hbar
\delta_{j}^{k}~, \label{3}%
\end{align}
one can construct a path-integral representation for the matrix elements
$G_{x}=\left\langle x_{out}\right\vert \hat{U}\left(  t_{out},t_{in}\right)
\left\vert x_{in}\right\rangle \,$ of the evolution operator $\hat{U}\left(
t,\acute{t}\right)  =\exp\left\{  -\frac{i}{\hbar}\hat{H}\left(  t-t^{\prime
}\right)  \right\}  $ in $x$-representation,%
\begin{equation}
G_{x}=\int Dp\int_{x_{(in)}-\theta p/2\hbar}^{x_{\left(  out\right)  }-\theta
p/2\hbar}Dq\exp\left\{  \frac{i}{\hbar}S^{\theta}\right\}  \,, \label{12}%
\end{equation}
where%
\begin{equation}
S^{\theta}=\int dt~\left[  p_{j}\dot{q}^{j}-H\left(  p,q\right)  -\dot{p}%
_{j}\theta^{ji}p_{i}/2\hbar\right]  ~, \label{11a}%
\end{equation}
see \cite{GitKu08}.

In quantum mechanics on commutative space, the action $S=\left.  S^{\theta
}\right\vert _{\theta=0}$ is just the Hamiltonian action of the classical
system under consideration. In the noncommutative case this action is modified
by the adding of a new term $\dot{p}_{k}\theta^{kj}p_{j}/2\hbar$. The action
(\ref{11a}) is called $\theta$-modified Hamiltonian action of classical
mechanics. As was mentioned in \cite{Deriglazov,Duval}, its quantization leads
exactly to the commutation relations (\ref{1}). Below, we demonstrate this
explicitly within the canonical quantization framework. To this end, let us
treat (\ref{11a}) as a Lagrangian action with generalized coordinates
$Q=\left(  q,p\right)  $ and Lagrange function $L=L\left(  Q,\dot{Q}\right)
,$
\begin{equation}
L=p_{j}\dot{q}^{j}-H\left(  p,q\right)  -\dot{p}_{j}\theta^{ji}p_{i}/2\hbar~.
\label{13}%
\end{equation}
Constructing the Hamiltonian formulation, we introduce the momenta%
\begin{equation}
\pi_{k}=\frac{\partial L}{\partial\dot{q}_{k}}=p_{k}\ ,\;\tilde{\pi}_{k}%
=\frac{\partial L}{\partial\dot{p}_{k}}=-\frac{\theta^{kj}}{2\hbar}p_{j}~,
\label{14}%
\end{equation}
and find the primary constraints to be $\Phi^{\left(  1\right)  }=\left(
\phi,\tilde{\phi}\right)  =0$,%
\begin{equation}
\phi_{k}=\pi_{k}-p_{k}\ ,\;\tilde{\phi}_{k}=\tilde{\pi}_{k}+\frac{\theta^{kj}%
}{2\hbar}p_{j}~. \label{15}%
\end{equation}
They are of second-class, $\det\left\{  \Phi^{\left(  1\right)  }%
,\Phi^{\left(  1\right)  }\right\}  \neq0.$ Performing a canonical
transformation to the new canonical variables (they are labeled by primes),
\begin{equation}
q_{k}^{\prime}=q_{k}\ ,\;p_{k}^{\prime}=p_{k}\ ,\;\pi_{k}^{\prime}=\pi
_{k}-p_{k}\ ,\;\tilde{\pi}_{k}^{\prime}=\tilde{\pi}_{k}-q_{k}~, \label{18}%
\end{equation}
we obtain the constraints of the special form:%
\begin{equation}
\pi_{k}^{\prime}=0\ ,\;q_{k}^{\prime}=-\tilde{\pi}_{k}^{\prime}-\frac
{\theta^{kj}}{2\hbar}p_{j}^{\prime}~,\nonumber
\end{equation}
see \cite{GitTy90}. Therefore, we can exclude $\pi_{k}^{\prime}=0$ and
$q_{k}^{\prime}$ from the consideration, in particular, from the Hamiltonian.
Then the commutation relations for the rest of the variables are canonical and
the new Hamiltonian reads as%

\begin{equation}
H^{\prime}\left(  q^{\prime},p^{\prime}\right)  =H\left(  -\tilde{\pi}%
_{k}^{\prime}-\frac{\theta^{kj}}{2\hbar}p_{j}^{\prime},p^{\prime}\right)
\ ,\;\left\{  p_{j}^{\prime},\tilde{\pi}_{k}^{\prime}\right\}  =\delta
_{jk}~.\nonumber
\end{equation}
Performing one more canonical transformation $p_{j}^{\prime}=p_{j}%
,\ \tilde{\pi}_{k}^{\prime}=-x_{k}$, we obtain%

\begin{equation}
H_{\theta}=H\left(  x-\frac{\theta\cdot p}{2\hbar},p\right)  ~,\ \left\{
x_{j},p_{k}\right\}  =\delta_{jk}~,\ \left\{  x_{j},x_{k}\right\}  =\left\{
p_{j},p_{k}\right\}  =0~. \label{20}%
\end{equation}

It follows from the action (\ref{13}) the following relations:%
\begin{equation}
\dot{p}_{i}=-\frac{\partial H\left(  q,p\right)  }{\partial q^{i}}\ ,\;\dot
{q}^{i}=\frac{\partial H\left(  q,p\right)  }{\partial p_{i}}+\frac
{\theta^{ij}}{\hbar}\frac{\partial H\left(  q,p\right)  }{\partial q^{j}}~.
\label{21a}%
\end{equation}
Using the relation (\ref{3}) we obtain the classical canonical relations for
the commutative coordinates, namely
\begin{equation}
\dot{p}_{i}=-\frac{\partial H_{\theta}}{\partial x^{i}}\ ,\;\dot{x}^{i}%
=\frac{\partial H_{\theta}}{\partial p_{i}}~. \label{21b}%
\end{equation}

Passing from (\ref{20}) to quantum theory, we obtain%

\begin{equation}
\hat{H}_{\theta}=H\left(  \hat{x}-\frac{\theta\hat{p}}{2\hbar},\hat{p}\right)
\ ,\;\left[  \hat{x}_{j},\hat{p}_{k}\right]  =i\hbar\delta_{jk}\ ,\;\left[
\hat{x}_{j},\hat{x}_{k}\right]  =0\ ,\;\left[  \hat{p}_{j},\hat{p}_{k}\right]
=0~, \label{21}%
\end{equation}
which corresponds to (\ref{3}).

\section{Charged particle in constant magnetic field on the noncommutative
plane}

\subsection{Classical motion}

Consider a classical nonrelativistic particle moving in the plane $\left(
x^{1},x^{2}\right)  $\ and interacting with a constant and uniform magnetic
field of intensity\textbf{ }$B$ perpendicular to the plane. Such a field can
be described by a vector potential $\mathbf{A}$ only ($A^{0}=0$). The
Hamiltonian of the particle is\footnote{The charge of an electron is $-e$,
with $e>0$.}%
\begin{equation}
H(\mathbf{x,p)}=\frac{1}{2m}\left[  \mathbf{p}+\frac{e}{c}\mathbf{A}\left(
\mathbf{x}\right)  \right]  ^{2},\ \mathbf{x}=\left(  x^{1},x^{2}\right)
,\;\mathbf{p}=\left(  p_{1},p_{2}\right)  ~. \label{bd44}%
\end{equation}

In what follows, we use alternatively the Landau gauge $\mathbf{A}_{L}$ and
the symmetric gauge $\mathbf{A}_{S}$,
\begin{align}
\mathbf{A}_{L}  &  =B\left(  0,x^{1}\right)  \ ,\;\mathbf{A}_{S}=\frac{1}%
{2}\left(  -Bx^{2},Bx^{1}\right)  \,\mathbf{,}\label{b.1}\\
\mathbf{A}_{L}  &  =\mathbf{A}_{S}+\boldsymbol{\nabla}f\ ,\;f=\frac{1}%
{2}Bx^{1}x^{2}. \label{b.2}%
\end{align}

Supposing that the minimal coupling is still valid, the corresponding
Hamiltonian $H\left(  p,q\right)  $ of (\ref{bd44}) for the non-commutative
variables $\mathbf{q}=\left(  q^{1},q^{2}\right)  $, which respect the algebra
(\ref{1}), can be constructed through the substitution $\mathbf{A}\left(
\mathbf{x}\right)  \rightarrow\mathbf{A}\left(  \mathbf{q}\right)  $,%
\begin{equation}
\mathbf{A}_{L}\left(  q\right)  =B\left(  0,q^{1}\right)  \ ,\;\mathbf{A}%
_{S}\left(  q\right)  =\frac{1}{2}\left(  -Bq^{2},Bq^{1}\right)  ~.
\label{b.1a}%
\end{equation}
In the case under consideration, we can set $\theta^{kj}=\theta\varepsilon
^{kj}$, $k,j=1,2,$ where $\varepsilon^{kj}$ is the Levi-Civita symbol,
$\varepsilon^{12}=1$, such that the classical equations of motion (\ref{21a}),
in the Landau gauge (\ref{b.1}), have the following solutions
\begin{align}
q^{1}  &  =q_{0}^{1}+R\cos\left(  \omega t+\phi\right)  \ ,\;\omega=\frac
{e}{cm}\left\vert B\right\vert ~,\nonumber\\
q^{2}  &  =q_{0}^{2}+\varepsilon R\sin\left(  \omega t+\phi\right)
\ ,\;\varepsilon=1+\frac{Be}{\hbar c}\theta~, \label{classic-landau}%
\end{align}
where $R$, $\phi$ and $q_{0}^{i}$\ are real constants. The motion is periodic,
with frequency $\omega$ (the cyclotron frequency), along an ellipse with
center $\left(  q_{0}^{1},q_{0}^{2}\right)  $ and eccentricity $\varepsilon$.
Hence, this elliptic deformation of the circle is due to the noncommutivity
parameter $\theta$ through $\varepsilon$. If we choose, instead of
(\ref{b.1}), the gauge $\mathbf{A}_{L}=-B(q^{2},0)$, then due to the
antisymmetry of $\theta^{jk}$ the axis of the ellipse will be interchanged.

In the symmetric gauge (\ref{b.2}) the solutions of the corresponding
classical equations of motion (\ref{21a}) have the form%
\begin{align}
&  q^{1}=q_{0}^{1}+R\cos\left(  \tilde{\omega}t+\phi\right)  \ ,\;q^{2}%
=q_{0}^{2}+R\sin\left(  \tilde{\omega}t+\phi\right)  ~,\nonumber\\
&  \tilde{\omega}=\omega\left\vert \mu_{S}\right\vert \ ,\;\mu_{S}=1-\frac
{eB}{4c\hbar}\theta~. \label{classic-sym}%
\end{align}
In this case the trajectory remains a circle with radius $R$ whereas the
non-commutativity modifies the frequency of motion $\tilde{\omega}$: the
latter differs from the cyclotron one $\omega$ by a factor that depends on the
algebraic value (in particular, on the direction) of the magnetic field. In
the case when $B\theta>0$, the frequency of oscillation decreases as
$\left\vert B\right\vert $ increases, and turns out to be zero for
$\theta=\theta_{c}^{S}=4c\hbar/eB.$

As was already mentioned in the literature \cite{DelDUGL07,IFUSP1648}, the
classical motion on the noncommutative plane is not gauge invariant under
gradient $U\left(  1\right)  $ gauge transformations of the external
electromagnetic field.

In both gauges, the relation between the minor radius $R$ of the ellipse and
the particle energy, defined as $E=H\left(  q,p\right)  $, reads as%
\[
\frac{E}{R^{2}}=\frac{m\omega^{2}}{2}~,
\]
which holds also in the commutative case. However, for $E=H_{\theta}\left(
x,p\right)  $, we have a $\theta$-dependent relation:%
\begin{equation}
\frac{E}{\tilde{R}^{2}}=\frac{m}{2}\left(  \mu_{S}\omega\right)
^{2}\ ,\;\tilde{R}^{2}=\left(  x^{1}\right)  ^{2}+\left(  x^{2}\right)  ^{2}~.
\label{raio}%
\end{equation}

\subsection{Quantum theory}

\paragraph{Symmetric gauge}

Passing to quantum theory, we choose first the symmetric gauge $\mathbf{A}%
_{S}$. Then%
\[
\hat{A}_{i}=-\frac{B}{2}\varepsilon_{ij}\left(  \hat{x}^{j}-\frac{\theta
}{2\hbar}\varepsilon^{jk}\hat{p}_{k}\right)  \ ,\;i,j,k=1,2~,
\]
and the quantum Hamiltonian (\ref{21}) takes the form:%
\begin{equation}
\hat{H}_{\theta}=\frac{1}{2\tilde{m}}\left(  \hat{P}_{1}^{2}+\hat{P}_{2}%
^{2}\right)  \ ,\;\tilde{m}=\frac{m}{\mu_{S}^{2}}\ ,\;\mu_{S}=1-\frac
{eB}{4c\hbar}\theta~. \label{r8}%
\end{equation}
Here $\hat{P}_{i},$ $i=1,2,$ are components of the kinematic momentum
operator,%
\begin{equation}
\hat{P}_{i}=\hat{p}_{i}-\frac{e\tilde{B}}{2c}\varepsilon_{ij}\hat{x}%
^{j}\ ,\;\left[  \hat{P}_{1},\hat{P}_{2}\right]  =-i\hbar\frac{e\tilde{B}}%
{c}\ ,\;\tilde{B}=\frac{B}{\mu^{S}}~. \label{km}%
\end{equation}

Like in the classical case, for $B\theta>0$, there exists a critical value
$\theta=\theta_{c}^{S}=4c\hbar/eB$, for which the Hamiltonian (\ref{r8}) does
not depend on the momenta,
\[
\hat{H}_{\theta_{c}^{S}}=\frac{1}{2m}\left(  \frac{eB}{2c}\right)  ^{2}\left[
\left(  \hat{x}^{1}\right)  ^{2}+\left(  \hat{x}^{2}\right)  ^{2}\right]  ~,
\]
and its eigenvectors describes localized states.

In the general case, (\ref{r8}) is a Hamiltonian of one-dimensional harmonic
oscillator with the spectrum%
\begin{equation}
E_{n}=\hbar\tilde{\omega}\left(  n+\frac{1}{2}\right)  \ ,\;\tilde{\omega
}=\frac{e}{c\tilde{m}}\left\vert \tilde{B}\right\vert =\frac{e}{cm}\left\vert
\mu_{S}B\right\vert \ ,\;n\in%
\mathbb{N}
\ . \label{r11}%
\end{equation}
The frequency $\tilde{\omega}$ coincide with the frequency of the classical
motion (\ref{classic-sym}) and, as was already mentioned, depends on the
algebraic value (in particular, on the direction) of the magnetic field.

It is convenient to introduce creation $\hat{a}^{+}$ and annihilation $\hat
{a}$\ operators,%
\begin{align}
&  \hat{a}=\frac{1}{\sqrt{2\tilde{m}\tilde{\omega}\hbar}}\left[  \frac
{e\tilde{B}}{2c}\hat{z}+2i\hat{p}_{z^{\ast}}\right]  \ ,\;\hat{a}^{+}=\frac
{1}{\sqrt{2\tilde{m}\tilde{\omega}\hbar}}\left[  \frac{e\tilde{B}}{2c}\hat
{z}^{\ast}-2i\hat{p}_{z}\right]  ~,\nonumber\\
&  \hat{z}=\hat{x}^{1}-i\hat{x}^{2}~,\ \hat{p}_{z}=-i\hbar\frac{\partial
}{\partial z}=\hat{p}_{z^{\ast}}^{+}\ ,\;\left[  \hat{a},\hat{a}^{+}\right]
=1~. \label{r6}%
\end{align}
In terms of such operators the Hamiltonian (\ref{r8}) assumes the form%
\[
\hat{H}_{\theta}=\hbar\tilde{\omega}\left(  \hat{N}+\frac{1}{2}\right)
\ ,\;\hat{N}=\hat{a}^{+}\hat{a}~.
\]

There exist additional operators of creation and annihilation,%
\begin{align}
&  \hat{b}=\frac{1}{\sqrt{2\tilde{m}\tilde{\omega}\hbar}}\left[  \frac
{e\tilde{B}}{2c}\hat{z}^{\ast}+2i\hat{p}_{z}\right]  \ ,\;\hat{b}^{+}=\frac
{1}{\sqrt{2\tilde{m}\tilde{\omega}\hbar}}\left[  \frac{e\tilde{B}}{2c}\hat
{z}-2i\hat{p}_{z^{\ast}}\right]  ~,\nonumber\\
&  \left[  \hat{b},\hat{b}^{+}\right]  =1~. \label{r9}%
\end{align}
They commute with $\hat{a}^{+}$, $\hat{a}$, and $\hat{H}_{\theta}$, such that
it is an integral of motion,%
\[
\left[  \hat{b},\hat{a}\right]  =\left[  \hat{b},\hat{a}^{+}\right]  =\left[
\hat{b},\hat{H}_{\theta}\right]  =0~.
\]
Thus, the eigenvectors of the Hamiltonian $\hat{H}_{\theta}$ corresponding to
the eigenvalues $E_{n}$, are%
\begin{equation}
\Psi_{mn}\left(  z,z^{\ast}\right)  =\frac{1}{\sqrt{m!}}\frac{1}{\sqrt{n!}%
}\left(  \hat{b}^{+}\right)  ^{m}\left(  \hat{a}^{+}\right)  ^{n}\psi
_{0}\left(  z,z^{\ast}\right)  ~, \label{r10}%
\end{equation}
with the ground state $\psi_{0}$ given by%
\begin{equation}
\hat{a}\psi_{0}=0\Longrightarrow\frac{\partial\psi_{0}}{\partial z^{\ast}%
}=-\frac{e\tilde{B}}{4}z\psi_{0}\Longrightarrow\psi_{0}=N\exp\left[
-\frac{e\tilde{B}}{8\hbar}\left\vert z\right\vert ^{2}\right]  ~, \label{r7}%
\end{equation}
where $N$ is a normalization factor. The above eigenfunctions are infinitely degenerate.

\paragraph{Landau gauge}

In the Landau gauge $\mathbf{A}_{L}$, the quantum Hamiltonian reads:%
\begin{align}
&  \hat{H}_{\theta}=\frac{\hat{p}_{1}^{2}}{2m}+\frac{1}{2\tilde{m}}\left(
\hat{p}_{2}+\frac{e\tilde{B}\hat{x}^{1}}{c}\right)  ^{2},\nonumber\\
&  \tilde{m}=\frac{m}{\mu^{2}},\ \tilde{B}=\frac{B}{\mu},\ \mu=1-\frac
{eB}{2c\hbar}\theta. \label{b.7}%
\end{align}
Here we obtain a different critical value $\theta=2c\hbar/eB=2\theta_{c}^{S}$ .

Introducing the kinematic momentum operators $\hat{P}_{1}$, and $\hat{P}_{2},$%
\[
\hat{P}_{1}=\hat{p}_{1}\ ,\;\hat{P}_{2}=\hat{p}_{2}+\frac{e\tilde{B}}{c}%
\hat{x}^{1}\Longrightarrow\left[  \hat{P}_{1},\hat{P}_{2}\right]
=-i\hbar\frac{e\tilde{B}}{c}~,
\]
we write (\ref{b.7}) as a one-dimensional harmonic oscillator Hamiltonian
\begin{align}
&  \hat{H}_{\theta}=\frac{1}{2m}\left[  \hat{P}_{1}^{2}+\left(  m\omega\hat
{Q}\right)  ^{2}\right]  \ ,\;\hat{P}\equiv\hat{P}_{1}\ ,\;\hat{Q}\equiv
\frac{c}{e\tilde{B}}\hat{P}_{2}~,\nonumber\\
&  \left[  \hat{Q},\hat{P}\right]  =i\hbar\ ,\;\omega=\frac{e}{mc}\left\vert
B\right\vert ~, \label{b.8}%
\end{align}
with cyclotron frequency $\omega$. Thus, in the Landau gauge, the spectrum is%
\[
E_{n}=\frac{eB}{mc}\left(  n+\frac{1}{2}\right)  \ ,\;n\in%
\mathbb{N}
\ .
\]
In this case, the annihilation operator $\hat{a}$ from (\ref{r6}) is commonly
changed into%
\begin{equation}
\hat{a}=\frac{1}{\sqrt{2m\omega\hbar}}\left[  \mu\hat{P}_{2}+i\hat{P}%
_{1}\right]  ~, \label{r6a}%
\end{equation}
such that%
\begin{equation}
\hat{H}_{\theta}=\hbar\omega\left(  \hat{a}^{+}\hat{a}+\frac{1}{2}\right)  ~.
\label{b.9}%
\end{equation}
In the gauge under consideration, $\hat{p}_{2}$ is a integral of motion, it
commutes with $\hat{H}_{\theta}$. Here the eigenvectors of $\hat{H}_{\theta}$
can be chosen as%

\[
\Psi_{n,k_{2}}\left(  z,z^{\ast}\right)  =\frac{1}{\sqrt{n!}}\left(  \hat
{a}^{+}\right)  ^{n}\psi_{0}\left(  x^{1}\right)  \psi_{k_{2}}\left(
x^{2}\right)  ~,
\]
with the state $\psi_{k2}$ and the ground state $\psi_{0}$\ given by%
\begin{align}
&  \hat{p}_{2}\psi_{k_{2}}=k_{2}\psi_{k_{2}}\Longrightarrow\psi_{k_{2}}\left(
x^{2}\right)  =\exp\left(  ik_{2}x^{2}\right)  ~,\nonumber\\
&  \hat{a}\psi_{0}=0\Longrightarrow\psi_{0}\left(  x^{1}\right)  =N\exp\left[
-\frac{\omega}{2\hbar}\left(  \frac{\mu}{\omega}k_{2}+x^{1}\right)
^{2}\right]  ~, \label{r7a}%
\end{align}
where $N$ is a normalization factor.

\section{Malkin-Man'ko coherent states on noncommutative plane}

The CS of a charged particle in a uniform magnetic filed were originally
constructed by Malkin and Man'ko \cite{MalMa68}. In fact, due to the double
analytic structure of the phase space, those states are the tensor products of
standard CS. In the general case, the phase space is $\mathbb{C}^{2}=\left\{
\mathbf{x}=\left(  \alpha,\beta\right)  \,,\,\alpha\in\mathbb{C}\,,\,\beta
\in\mathbb{C}\right\}  $ and the realization of such a space can be
constructed as the Hilbert space
\[
L^{2}\left(  \mathbb{C}^{2},\mu\left(  d\mathbf{x}\right)  \right)
=L^{2}\left(  \mathbb{C},\mu\left(  d\alpha\right)  \right)  \otimes
L^{2}\left(  \mathbb{C},\mathbb{C},\mu\left(  d\beta\right)  \right)  ~,
\]
provided with adequate measures $\mu\left(  d\alpha\right)  $ and $\mu\left(
d\beta\right)  $. For the specific case of the standard CS, the measure is
chosen to be%
\begin{equation}
\mu\left(  d\mathbf{x}\right)  =e^{-\left\vert \alpha\right\vert ^{2}}%
\,\frac{d^{2}\alpha}{\pi}\,e^{-\left\vert \beta\right\vert ^{2}}\,\frac
{d^{2}\beta}{\pi}\,, \label{c5}%
\end{equation}
where $d^{2}\alpha$ and $d^{2}\beta$ are the respective Lebesgue measures on
the complex planes. In this Hilbert space we can define the following
orthonormal set of functions
\begin{equation}
\Phi_{m,n}\left(  \mathbf{x}\right)  \equiv\frac{\bar{\alpha}^{m}}{\sqrt{m!}%
}\,\frac{\bar{\beta}^{n}}{\sqrt{n!}}\,, \label{c4}%
\end{equation}
that we put in one-to-one correspondence with the elements $|m,n\rangle$,
$m,n\in\mathbb{N}$, of any orthonormal basis of a separable Hilbert space
$\mathcal{H}$. We now introduce the CS corresponding to this choice of
orthonormal set. They are elements of $\mathcal{H}$ defined by
\begin{equation}
\left\vert \alpha,\beta\right\rangle =\left\vert \alpha\right\rangle
\otimes\left\vert \beta\right\rangle \equiv\exp\left(  -\frac{\left\vert
\alpha\right\vert ^{2}+\left\vert \beta\right\vert ^{2}}{2}\right)  \sum
_{m,n}\frac{\alpha^{m}}{\sqrt{m!}}\,\frac{\beta^{n}}{\sqrt{n!}}\left\vert
m,n\right\rangle \,. \label{c2}%
\end{equation}
By construction, these normalized states are labeled by points of
$\mathbb{C}^{2}$ and form a continuous overcomplete set resolving the unity in
$\mathcal{H}$. In the case of the symmetric gauge $\mathbf{A}_{S}$
(\ref{b.1a}), the standard CS are constructed by choosing the states
(\ref{r10}), eigenstates of the Hamiltonian $H_{\theta}$ (\ref{r8}), as an
orthonormal basis. With this choice, (\ref{c2}) can be written as%
\begin{equation}
\left\vert \alpha,\beta\right\rangle =\hat{Z}\left\vert 0,0\right\rangle
\,,\ \hat{Z}=\exp\left(  \alpha\hat{a}^{+}-\alpha^{\ast}\hat{a}\right)
\exp\left(  \beta\hat{b}^{+}-\beta^{\ast}\hat{b}\right)  ~, \label{c0}%
\end{equation}
where the operators $\hat{a}$ and $\hat{b}$ are given by (\ref{r6}) and
(\ref{r9}), and the ground state $\Psi_{00}$ (\ref{r10}), supposed normalized,
is denoted here by $\left\vert 0,0\right\rangle $. The above CS, which we have
written using displacement operator \cite{KlaSk85}, correspond to the CS
obtained by Malkin and Man'ko.

The time evolution of these coherent states can be obtained as%
\begin{equation}
\left\vert \alpha,\beta;t\right\rangle =\exp\left(  -i\frac{\hat{H}_{\theta}%
}{\hbar}t\right)  \left\vert \alpha,\beta\right\rangle =\exp\left(
-i\frac{\tilde{\omega}}{2}t\right)  \left\vert \alpha\exp\left(
-i\tilde{\omega}t\right)  ,\beta\right\rangle ~. \label{cs1}%
\end{equation}
We note here that the Hamiltonian $\hat{H}_{\theta}$ acts as the identity on
the $\left\vert \beta\right\rangle $ part of the CS, $\hat{H}_{\theta}%
=\hbar\tilde{\omega}\left(  \hat{N}_{a}+1/2\right)  \otimes I$.\ As we will
see, these states describe circular trajectories and the parameter $\alpha$ is
related to the radius of the orbit, while $\beta$ is related to the center of
the orbit.

For the Landau gauge one can define the semi-coherent states%
\begin{align}
\left\vert \alpha,k_{2}\right\rangle  &  =\exp\left[  -\frac{1}{2}\left\vert
\alpha\right\vert ^{2}\right]  \exp\left(  \alpha\hat{a}^{+}\right)
\left\vert 0,k_{2}\right\rangle ~,\nonumber\\
\left\vert 0,k_{2}\right\rangle  &  =\left\vert 0\right\rangle \otimes
\left\vert k_{2}\right\rangle ~,\ \hat{a}\left\vert 0\right\rangle
=0~,\ \hat{p}_{2}\left\vert k_{2}\right\rangle =k_{2}\left\vert k_{2}%
\right\rangle ~, \label{cs2}%
\end{align}
where the operators $\hat{a}$ is now given by (\ref{r6a}) and the ground state
$\psi_{0}$, denoted here $\left\vert 0\right\rangle $, is given by
(\ref{r7a}). Note that we could as well define Malkin-Man'ko coherent states
for this case.

Let us study the mean value evolution of the coordinate operators for the
above CS. For the symmetric gauge let us use the kinematical momentum operator
(\ref{km}) and introduce the centre-coordinate operator \cite{JohLi49}%
\begin{equation}
\hat{x}_{0}^{i}=\hat{x}^{i}-\frac{1}{\tilde{m}\tilde{\omega}}\varepsilon
^{ij}\hat{P}_{j}\ ,\;\tilde{m}\tilde{\omega}=\frac{m\omega}{\mu_{S}}%
\ ,\;\mu_{S}=1-\frac{eB}{4c\hbar}\theta\ ,\;i,j=1,2~, \label{k1}%
\end{equation}
which are integral of motion%
\[
\left[  \hat{x}_{0}^{1},\hat{H}_{\theta}\right]  =\left[  \hat{x}_{0}^{2}%
,\hat{H}_{\theta}\right]  =0\ ,\;\left[  \hat{x}_{0}^{1},\hat{x}_{0}%
^{2}\right]  =\frac{i\hbar}{\tilde{m}\tilde{\omega}}~.
\]
From (\ref{k1}) and (\ref{r6}) we have%
\[
\hat{x}^{1}-\hat{x}_{0}^{1}=\sqrt{\frac{\hbar}{2\tilde{m}\tilde{\omega}}%
}\left(  \hat{a}+\hat{a}^{+}\right)  \ ,\ \hat{x}^{2}-\hat{x}_{0}^{2}%
=i\sqrt{\frac{\hbar}{2\tilde{m}\tilde{\omega}}}\left(  \hat{a}-\hat{a}%
^{+}\right)  ~.
\]
From the fact that $\left\vert \alpha,\beta\right\rangle $ is an eigenvector
of $\hat{a}$ with eigenvalue $\alpha$, we easily derive:%
\begin{align*}
&  \left\langle \alpha,\beta\right\vert \left(  \hat{x}^{1}-\hat{x}_{0}%
^{1}\right)  \left\vert \alpha,\beta\right\rangle =\sqrt{\frac{2\hbar}%
{\tilde{m}\tilde{\omega}}}\operatorname{Re}\left(  \alpha\right)  ~,\\
&  \left\langle \alpha,\beta\right\vert \left(  \hat{x}^{2}-\hat{x}_{0}%
^{2}\right)  \left\vert \alpha\beta\right\rangle =-\sqrt{\frac{2\hbar}%
{\tilde{m}\tilde{\omega}}}\operatorname{Im}\left(  \alpha\right)  ~.
\end{align*}
Writing $\alpha=\hbar^{-1/2}R~e^{-i\phi}$, with $R$ and $\phi$ real constants,
and using the time-dependence of $\left\vert \alpha,\beta;t\right\rangle $
(\ref{cs1})\ we finally get%
\begin{align*}
\left\langle \alpha,\beta;t\right\vert \left(  \hat{x}^{1}-\hat{x}_{0}%
^{1}\right)  \left\vert \alpha,\beta;t\right\rangle  &  =\sqrt{\frac{2}%
{\tilde{m}\tilde{\omega}}}R\cos\left(  \tilde{\omega}t+\phi\right)  ~,\\
\left\langle \alpha,\beta;t\right\vert \left(  \hat{x}^{2}-\hat{x}_{0}%
^{2}\right)  \left\vert \alpha,\beta;t\right\rangle  &  =\sqrt{\frac{2}%
{\tilde{m}\tilde{\omega}}}R\sin\left(  \tilde{\omega}t+\phi\right)  ~,
\end{align*}
where the frequency of oscillation $\tilde{\omega}$\ concords with the
classical expression (\ref{classic-sym}).

Alternatively, for the Landau gauge we have from (\ref{cs2}),%
\[
\left\langle \alpha,k_{2};t\right\vert \hat{x}^{1}\left\vert \alpha
,k_{2};t\right\rangle =-\sqrt{\frac{m\hbar}{2\omega}}\operatorname{Re}\left(
\alpha\right)  -\frac{\mu}{\omega}k_{2}\ ,\;\left\langle \alpha,k_{2}%
\right\vert \hat{p}_{1}\left\vert \alpha,k_{2}\right\rangle =\sqrt
{\frac{m\omega\hbar}{2}}\operatorname{Im}\left(  \alpha\right)  ~,
\]
with $\mu$ given by (\ref{b.8}). Writing $\alpha=\hbar^{-1/2}R~e^{-i\phi}$,
with $R$ and $\phi$ real constants, and using the time-dependence of
$\left\vert \alpha\right\rangle $ (\ref{cs1})\ we get%
\[
\left\langle \alpha,k_{2}\right\vert \hat{x}^{1}\left\vert \alpha
,k_{2}\right\rangle =-\sqrt{\frac{m}{2\omega}}2R\cos\left(  \omega
t+\phi\right)  -\frac{\mu}{\omega}k_{2}~.
\]
Once again, the frequency of oscillation $\omega$\ concords with the classical
expression (\ref{classic-landau}).

As expected, by computing the mean values $\left\langle \hat{f}\right\rangle
=\left\langle \alpha, \beta\right\vert \hat{f}\left\vert \alpha,
\beta\right\rangle $\ and the dispersion $\left(  \Delta\hat{f}\right)
^{2}=\left\langle \hat{f}^{2}\right\rangle -\left\langle \hat{f}\right\rangle
^{2}$, we see that the above CS saturate the uncertainty relations,%

\[
\Delta\hat{x}_{i}=\sqrt{\frac{\mu c\hbar}{2B\left\vert e\right\vert }%
}\ ,\;\Delta\hat{p}_{i}=\sqrt{\frac{\hbar B\left\vert e\right\vert }{2c\mu}%
}\ ,\;\Delta\hat{x}_{i}\Delta\hat{p}_{i}=\frac{\hbar}{2}~,
\]
where $i=1,2$ and $\mu=\mu_{S}$ is given by (\ref{r8}) for the symmetric gauge
and for the Landau gauge (the semi-coherent states) $i=1$ and $\mu$ is given
by (\ref{b.8}).

We recall that standard CS $\left\vert \alpha,\beta\right\rangle $
(\ref{c0})\ resolve the unity operator in the Hilbert space spanned by the
eigenfunctions $\left\vert m,n\right\rangle $:%
\[
\int_{%
\mathbb{C}
}\int_{%
\mathbb{C}
}\frac{d^{2}\alpha~d^{2}\beta}{\pi^{2}}\left\vert \alpha,\beta\right\rangle
\left\langle \alpha,\beta\right\vert =I~.
\]
Hence, they allow a quantization \textit{\`{a} la} \textquotedblleft
Berezin-Klauder" or \textquotedblleft anti-Wick" quantization \cite{Ber71} of
the classical quantities $f\left(  \alpha,\beta\right)  $ through the
correspondence
\begin{equation}
f\left(  \zeta,\bar{\zeta}\right)  \mapsto\int_{%
\mathbb{C}
}\int_{%
\mathbb{C}
}\frac{d^{2}\alpha~d^{2}\beta}{\pi^{2}}f\left(  \alpha,\beta\right)
\left\vert \alpha,\beta\right\rangle \left\langle \alpha,\beta\right\vert
\equiv\hat{f}~. \label{c3}%
\end{equation}
For mild conditions on $f$, this linear map produces a well-defined operator
$\hat{f}$ in $\mathcal{H}$. Starting from the classical quantities $\left(
x^{1},x^{2}\right)  $, with $z=x^{1}-ix^{2}$, using the explicit form of
$\hat{a}$ and $\hat{b}$ ($\hat{a}^{+}$ and $\hat{b}^{+}$) in (\ref{r6}), and
the fact that $\alpha$ and $\beta$ ($\alpha^{\ast}$ and $\beta^{\ast}$)\ are
their respective eigenvalues, we see that these complex parameters are related
with the classical quantities\ by%
\begin{align}
x^{1}  &  =\sqrt{\frac{2\hbar}{\tilde{\omega}\tilde{m}}}\left(
\operatorname{Re}\alpha+\operatorname{Re}\beta\right)  \ ,\;x^{2}=\sqrt
{\frac{2\hbar}{\tilde{\omega}\tilde{m}}}\left(  \operatorname{Im}%
\beta-\operatorname{Im}\alpha\right)  ~,\nonumber\\
p_{1}  &  =\sqrt{\frac{\tilde{m}\tilde{\omega}\hbar}{2}}\left(
\operatorname{Im}\alpha+\operatorname{Im}\beta\right)  \ ,\;p_{2}=\sqrt
{\frac{\tilde{m}\tilde{\omega}\hbar}{2}}\left(  \operatorname{Re}%
\alpha-\operatorname{Re}\beta\right)  ~. \label{c1}%
\end{align}
Using the decomposition (\ref{c2})\ it is a matter of simple calculation to
prove that (\ref{c3}),\ for $f\left(  \alpha,\beta\right)  =\alpha$ and
$f\left(  \alpha,\beta\right)  =\beta$, give%
\[
\alpha\mapsto\int_{%
\mathbb{C}
}\int_{%
\mathbb{C}
}\frac{d^{2}\alpha\ d^{2}\beta}{\pi^{2}}\ \alpha\left\vert \alpha
,\beta\right\rangle \left\langle \alpha,\beta\right\vert =\hat{a}%
\ ,\;\beta\mapsto\int_{%
\mathbb{C}
}\int_{%
\mathbb{C}
}\frac{d^{2}\alpha\ d^{2}\beta}{\pi^{2}}\ \beta\left\vert \alpha
,\beta\right\rangle \left\langle \alpha,\beta\right\vert =\hat{b}~,
\]
and, in the same way, we have $\alpha^{\ast}\mapsto\hat{a}^{+}~,\ \beta^{\ast
}\mapsto\hat{b}^{+}$. The above expressions allows perform the CS quantization
of the classical quantities (\ref{c1}),%
\begin{align}
x^{1}  &  \mapsto\hat{x}^{1}=\sqrt{\frac{2\hbar}{\tilde{\omega}\tilde{m}}%
}\left(  \hat{a}+\hat{a}^{+}+\hat{b}+\hat{b}^{+}\right)  \ ,\;p_{1}\mapsto
\hat{p}_{1}=\frac{i}{2}\sqrt{\frac{\tilde{m}\tilde{\omega}\hbar}{2}}\left(
\hat{a}^{+}-\hat{a}-\hat{b}+\hat{b}^{+}\right)  ~,\nonumber\\
x^{2}  &  \mapsto\hat{x}^{2}=i\sqrt{\frac{\hbar}{2\tilde{\omega}\tilde{m}}%
}\left(  \hat{a}-\hat{a}^{+}-\hat{b}+\hat{b}^{+}\right)  \ ,\;p_{2}\mapsto
\hat{p}_{2}=\sqrt{\frac{\tilde{m}\tilde{\omega}\hbar}{2}}\left(  \hat{a}%
+\hat{a}^{+}-\hat{b}-\hat{b}^{+}\right)  ~. \label{c6}%
\end{align}
Using the above relations we can see explicitly that the definition of a
classical variable\ in the form (\ref{3}),%

\[
q^{k}=x^{k}-\frac{1}{2\hbar}\theta\varepsilon^{kj}p_{j}\mapsto\hat{x}%
^{k}-\frac{1}{2\hbar}\theta\varepsilon^{kj}\hat{p}_{j}~,
\]
reproduces the adequate quantum theory with non-commuting coordinates for the
position operator where $\left[  \hat{q}^{1},\hat{q}^{2}\right]  =i\theta$.

\section{Circular-coherent states on the noncommutative plane}

In this section we construct a different CS family for the problem of a
charged particle in a non-commutative plane. Due to the nature of the behavior
of a charged particle in a uniform magnetic field, our approach will make use
of the coherent states for the motion of a quantum particle on a circle. In
the work \cite{KowRe05} the authors propose the construction of CS for a
particle in a uniform magnetic field by precisely using the CS for the circle.
The latter are constructed from the angular momentum operator $\hat{J}$ and
the unitary operator $\hat{U}$ that represents the position of the particle on
the unit circle. These operators obey the commutation relations
\cite{KowReP96},%
\begin{equation}
\left[  \hat{J},\hat{U}\right]  =U\ ,\;\left[  \hat{J},\hat{U}^{+}\right]
=-\hat{U}^{+}~. \label{k0}%
\end{equation}
The introduction of these CS permits to avoid the problem of the infinite
degeneracy present in the approach followed by Man'ko and Malkin, and, in
addition, takes into in account the momentum part of the phase space.
Consequently, the so obtained CS offer a better way to compare the quantum
behavior of the system with the classical trajectories in the phase space. In
the present approach, we give a generalization, or a squeezed version, of
these circular CS for the particle in the magnetic field and on a
noncommutative plane.

In the symmetric gauge, let us introduce the centre-coordinate, as in
(\ref{k1}), operators
\begin{equation}
\hat{x}_{0}^{1}=\hat{x}^{1}-\frac{1}{\tilde{m}\tilde{\omega}}\hat{P}%
_{2}\,,\;\hat{x}_{0}^{2}=\hat{x}^{2}+\frac{1}{\tilde{m}\tilde{\omega}}\hat
{P}_{1}\,, \label{centcoord}%
\end{equation}
which are integral of motion, $[H_{\tilde{\theta}},\hat{x}_{0}^{i}]=0$, and
the relative motion coordinates,
\begin{equation}
\hat{r}^{1}=\hat{x}^{1}-\hat{x}_{0}^{1}=\frac{1}{\tilde{m}\tilde{\omega}}%
\hat{P}_{2}\,,\;\hat{r}^{2}=\hat{x}^{2}-\hat{x}_{0}^{1}=-\frac{1}{\tilde
{m}\tilde{\omega}}\hat{P}_{1}\,. \label{relcoord}%
\end{equation}
Next let us define the operators
\begin{equation}
\hat{r}_{0\pm}=\hat{x}_{0}^{1}\pm i\hat{x}_{0}^{2}\,,\;\hat{r}_{\pm}=\hat
{r}^{1}\pm i\hat{r}^{2}=\frac{1}{\tilde{m}\tilde{\omega}}\left(  \hat{P}%
_{2}\mp i\hat{P}_{1}\right)  \,. \label{rpm}%
\end{equation}
They obey the commutation rules
\begin{equation}
\left[  \hat{r}_{0+},\hat{r}_{0-}\right]  =2\frac{\hbar}{\tilde{m}%
\tilde{\omega}}\,,\;\left[  \hat{r}_{+},\hat{r}_{-}\right]  =-2\frac{\hbar
}{\tilde{m}\tilde{\omega}}\,,\;\left[  \hat{r}_{0\pm},\hat{r}_{\pm}\right]
=0\,. \label{comruler}%
\end{equation}
We now define the angular momentum operator $\hat{J}$, which is just
proportional to the Hamiltonian (\ref{r8}),%
\[
\hat{J}=\hat{r}_{1}\hat{P}_{2}-\hat{r}_{2}\hat{P}_{1}=\frac{2}{\tilde{\omega}%
}\hat{H}_{\theta}=\tilde{m}\tilde{\omega}\hat{r}_{+}\hat{r}_{-}+\hbar
=\tilde{m}\tilde{\omega}\hat{r}_{-}\hat{r}_{+}-\hbar\,.
\]
The above expression coincides with the classical one (\ref{raio}). Due to the
rules,
\begin{equation}
\left[  J,\hat{r}_{0\pm}\right]  =0\,,\;\left[  J,\hat{r}_{\pm}\right]
=\pm2\hbar\hat{r}_{\pm}\,, \label{rotcomrul}%
\end{equation}
the operator $\hat{J}$ can be identified as the generator of rotations about
the axis passing through the classical point $(x_{0}^{1},x_{0}^{2})$ and
perpendicular to the $(x^{1},x^{2})$ plane. The nonunitary operator $\hat
{r}_{-}$ is the counterpart of the unitary operator $\hat{U}$ in (\ref{k0}),
which describes to a certain extent the angular position of the particle on a
circle. Actually, the factorization of $\hat{r}_{-}$where, in the present
case,%
\[
\hat{r}_{-}=\hat{R}_{-}\hat{V}~,\;\hat{R}_{-}=\sqrt{\hat{r}_{-}\hat{r}%
_{-}^{\dag}}%
\]
allows to view $\hat{V}$ as a unitary operator related to $\hat{U}$.

The symmetries and the integrability of the model can be encoded into the two
independent Weyl-Heisenberg algebras issued from the rules (\ref{rotcomrul}),
one for the center of the circular orbit and the other for the relative
motion. They allow one to construct the Fock space\ (\ref{r10}), with
orthonormal basis $\{|m,n\rangle\equiv|m\rangle\otimes|n\rangle\,,\,m,n\in
\mathbb{N}\}$, as repeated actions of the raising operators $\hat{r}_{0-}$ and
$\hat{r}_{+}$,
\begin{equation}
\hat{r}_{0-}|m\rangle=\sqrt{\frac{2\hbar(m+1)}{\tilde{m}\tilde{\omega}}%
}|m+1\rangle\,,\;\hat{r}_{+}|n\rangle=\sqrt{\frac{2\hbar(n+1)}{\tilde{m}%
\tilde{\omega}}}|n+1\rangle\,. \label{fockbasis+}%
\end{equation}
On the other hand, we have
\begin{equation}
\hat{r}_{0+}|m\rangle=\sqrt{\frac{2\hbar m}{\tilde{m}\tilde{\omega}}%
}|m-1\rangle\,,\;\hat{r}_{-}|n\rangle=\sqrt{\frac{2\hbar n}{\tilde{m}%
\tilde{\omega}}}|n-1\rangle\,, \label{fockbasis-}%
\end{equation}
and the eigenvalue equation
\begin{equation}
\hat{J}\left\vert m,n\right\rangle =\left(  2n+1\right)  \hbar\left\vert
m,n\right\rangle ~. \label{k11}%
\end{equation}

The circular CS $\left\vert z_{0},\zeta\right\rangle $, as they were proposed
in \cite{KowRe05} (although with some notational differences), are constructed
in the Hilbert space spanned by the orthonormal basis $\{|m,n\rangle\}$ as
solutions to the eigenvalue equations:
\begin{equation}
\hat{r}_{0+}\left\vert z_{0},\zeta\right\rangle =z_{0}\left\vert z_{0}%
,\zeta\right\rangle \ ,\;\hat{Z}\left\vert z_{0},\zeta\right\rangle
=\zeta\left\vert z_{0},\zeta\right\rangle ~,\;z_{0},\zeta\in\mathbb{C}\,,
\label{A.1}%
\end{equation}
where the operator $\hat{Z}$ is defined by
\begin{equation}
\hat{Z}=e^{\frac{1}{2}(\hat{J}/\hbar+1)}\hat{r}_{-}~. \label{opZ}%
\end{equation}

The projection of the CS (\ref{A.1})\ in this Fock basis\ reads as
\begin{align*}
&  \left\langle m,n\right\vert \left.  \zeta,z_{0}\right\rangle =\frac
{e^{-\frac{\left\vert \tilde{z}_{0}\right\vert ^{2}}{2}}}{\sqrt{\mathcal{E}%
(|\tilde{\zeta}|^{2})}}\frac{\tilde{z}_{0}^{m}}{\sqrt{m!}}\,\frac{\tilde
{\zeta}^{n}}{\sqrt{n!}}e^{-\frac{1}{2}n(n+1)}~,\\
&  \tilde{m}\tilde{\omega}=e\left\vert \tilde{B}\right\vert \frac{\hbar}%
{c}\ ,\;\tilde{B}=\frac{B}{\mu_{S}}~,\;\mu_{S}=1-\frac{eB}{4c\hbar}\theta~,
\end{align*}
where, for notational convenience, we have introduced the dimensionless
variables
\begin{equation}
\tilde{z}_{0}=\sqrt{\frac{\tilde{m}\tilde{\omega}}{2\hbar}}\,z_{0}%
\,,\quad\tilde{\zeta}=\sqrt{\frac{\tilde{m}\tilde{\omega}}{2\hbar}}\,\zeta\,.
\label{nodimvar}%
\end{equation}
The normalization factor involves the function
\begin{equation}
\mathcal{E}\left(  t\right)  =\sum_{n=0}^{\infty}e^{-n(n+1)}\frac{t^{n}}%
{n!}\,. \label{nzeta}%
\end{equation}

We can generalize the above formulation and, consequently, obtain a squeezed
version of the CS (\ref{A.1}), defining the CS of the charge in a uniform
magnetic field as the eigenvector of the commuting operators $\hat{r}_{0+}$
and $\hat{Z}_{\lambda}$ \cite{GazBaG09},
\begin{equation}
\hat{r}_{0+}\left\vert z_{0},\zeta\right\rangle =z_{0}\left\vert z_{0}%
,\zeta\right\rangle \ ,\;\hat{Z}_{\lambda}\left\vert z_{0},\zeta\right\rangle
=\zeta\left\vert z_{0},\zeta\right\rangle ~, \label{k12}%
\end{equation}
where
\begin{equation}
\hat{Z}_{\lambda}=\exp\left[  \frac{\lambda}{4}\left(  \frac{\hat{J}}{\hbar
}+1\right)  \right]  \hat{r}_{-}=\sum_{n\geq1}e^{\frac{\lambda}{2}n}\sqrt
{n}|n-1\rangle\langle n|~. \label{zlambda}%
\end{equation}

The operator $\hat{Z}_{\lambda}$ coincides with $\hat{Z}$ from (\ref{A.1}) for
$\lambda=2$, and with just $\hat{r}_{-}$ for $\lambda=0$, i.e., the case in
which we have the tensor product of standard coherent states, called in this
context the Malkin-Man'ko CS \cite{MalMa68}. For an arbitrary $\lambda\geq0$,
$\hat{Z}_{\lambda}$ controls the dispersion relations of the angular momentum
$\hat{J}$ and of the \textquotedblleft position operator\textquotedblright%
\ $\hat{r}_{-}$. Note the expressions in terms of the number operator $\hat
{N}$ and the resulting commutation rule,
\begin{align}
&  \hat{Z}_{\lambda}\hat{Z}_{\lambda}^{\dag}=\frac{2\hbar}{\tilde{m}%
\tilde{\omega}}\partial_{\lambda}e^{\lambda(\hat{N}+1)}\,,\;\hat{Z}_{\lambda
}^{\dag}\hat{Z}_{\lambda}=\frac{2\hbar}{\tilde{m}\tilde{\omega}}%
\partial_{\lambda}e^{\lambda\hat{N}}\,,\nonumber\\
&  [\hat{Z}_{\lambda},\hat{Z}_{\lambda}^{\dag}]=2\frac{2\hbar}{\tilde{m}%
\tilde{\omega}}\partial_{\lambda}\left[  \sinh(\lambda/2)\,e^{\lambda(\hat
{N}+1/2)}\right]  \,. \label{comrulZ}%
\end{align}

We will call the states defined in (\ref{k12}) $\lambda$-coherent states
($\lambda$-CS). In the Fock basis (\ref{k11})\ these CS read as
\cite{GazBaG09}
\begin{equation}
\left\vert z_{0},\zeta\right\rangle =\frac{e^{-\frac{\left\vert \tilde{z}%
_{0}\right\vert ^{2}}{2}}}{\sqrt{\mathcal{E}_{\lambda}(|\tilde{\zeta}|^{2})}%
}\sum_{m,n}\frac{\tilde{z}_{0}^{m}}{\sqrt{m!}}\,\frac{\tilde{\zeta}^{n}}%
{\sqrt{n!}}e^{-\frac{\lambda}{4}n(n+1)}\left\vert m,n\right\rangle \,,
\label{csdef}%
\end{equation}
where, for a fixed value of $\lambda$ in (\ref{k12}), the normalization
function $\mathcal{E}_{\lambda}\left(  t\right)  $ is a kind of generalized
\textquotedblleft exponential\textquotedblright\
\begin{equation}
\mathcal{E}_{\lambda}\left(  t\right)  =\sum_{n=0}^{\infty}e^{-\frac{\lambda
n\left(  n+1\right)  }{2}}\frac{t^{n}}{n!}\equiv\sum_{n=0}^{\infty}\frac
{t^{n}}{x_{n}!}\,, \label{nzetal}%
\end{equation}
where%
\begin{equation}
x_{n}=e^{n\lambda}n=\partial_{\lambda}e^{n\lambda}\,,x_{n}!=x_{1}x_{2}\cdots
x_{n}\,,\quad x_{0}!=1\,. \label{genfact}%
\end{equation}
The complex numbers $z_{0}$ and $\zeta$ parameterize, respectively, the
position of the centre of the circle and the classical phase space of the
circular motion. As was shown in \cite{KowRe05} these CS have some properties
that made them more suitable to describe the classical behavior of a charged
particle in a magnetic field, in comparison with the Malkin-Man'ko CS
\cite{MalMa68}. Besides, as we show in the next section, our generalization
with the $\lambda$ parameter can be explored to improve in an appreciable way
these interesting characteristics.

The $\lambda$-CS $\left\vert z_{0},\zeta\right\rangle $ (\ref{k12}) are the
tensor product of the states $\left\vert z_{0}\right\rangle $ and $\left\vert
\zeta\right\rangle $, where the first one is the standard CS described in the
previous section. So, in order to perform the Berezin-Klauder-Toeplitz
quantization using our CS, we concentrate only on the states $\left\vert
\zeta\right\rangle $. For convenience, we put $\hbar=\tilde{m}\tilde{\omega
}/2=1$, and so $\tilde{\zeta}=\zeta$. Then, in the basis $\{|n\rangle\}$, the
latter admits the decomposition%
\begin{equation}
\left\vert \zeta\right\rangle =\frac{1}{\sqrt{\mathcal{E}_{\lambda}%
(|\zeta|^{2})}}\sum_{n=0}^{+\infty}\frac{\zeta^{n}}{\sqrt{x_{n}!}}\left\vert
n\right\rangle ~,\quad x_{n}=e^{n\lambda}\,n\,. \label{zetacs}%
\end{equation}

As was shown in \cite{GazBaG09} the CS states $\left\vert \zeta\right\rangle
$\ resolve the unity operator in the Hilbert space spanned by the kets
$\left\vert n\right\rangle $,%
\[
\int_{\mathbb{C}}\,\varpi_{\lambda}\left(  \left\vert \zeta\right\vert
^{2}\right)  \,\frac{d^{2}\zeta}{\pi}~\mathcal{E}_{\lambda}(|\zeta
|^{2})\,\left\vert \zeta\right\rangle \left\langle \zeta\right\vert =I~,
\]
where the weight function $\varpi_{\lambda}$ is given under the form of the
Laplace transform,%
\begin{equation}
\varpi_{\lambda}\left(  t\right)  =\frac{e^{-\lambda/2}}{\sqrt{2\pi\lambda}%
}\int_{0}^{+\infty}du\,\exp\left(  -e^{-\lambda/2}tu\right)  e^{-\frac{\left(
\ln u\right)  ^{2}}{2\lambda}}=\frac{e^{-\lambda/2}}{\sqrt{2\pi\lambda}%
}\,\mathcal{L}\left[  e^{-\frac{\left(  \ln u\right)  ^{2}}{2\lambda}}\right]
\left(  e^{-\lambda/2}t\right)  \,.\nonumber
\end{equation}
The corresponding CS quantization of functions on the complex plane, the phase
space for the relative motion, is the map
\begin{equation}
f\left(  \zeta,\bar{\zeta}\right)  \mapsto\int_{\mathbb{C}}\frac{d^{2}\zeta
}{\pi}\,\varpi_{\lambda}\left(  \left\vert \zeta\right\vert ^{2}\right)
\,f\left(  \zeta,\bar{\zeta}\right)  \,\mathcal{E}_{\lambda}\left(  \left\vert
\zeta\right\vert ^{2}\right)  \,\left\vert \zeta\right\rangle \left\langle
\zeta\right\vert \overset{\mathrm{def}}{=}\hat{f}\,. \label{qcsquant}%
\end{equation}

Using the fact that the weight function $\varpi_{\lambda}$\ solves the
following moment problem \cite{GazBaG09},
\[
\int_{0}^{\infty}t^{n}\varpi_{\lambda}\left(  t\right)  \,dt=n!\,\exp\left\{
\dfrac{\lambda n\left(  n+1\right)  }{2}\right\}  \,,\ \lambda\geq0\,,
\]
it is easy to obtain the quantization of the variable $\zeta$,%
\begin{align}
\zeta\mapsto\hat{\zeta}=  &  \int_{\mathbb{C}}\frac{d^{2}\zeta}{\pi^{2}}%
\varpi_{\lambda}\left(  \left\vert \zeta\right\vert ^{2}\right)
\,\mathcal{E}_{\lambda}\left(  \left\vert \zeta\right\vert ^{2}\right)
\,\zeta~\left\vert \zeta\right\rangle \left\langle \zeta\right\vert
\nonumber\\
=  &  \sum_{n}\exp\left[  \frac{\lambda}{2}n\right]  \sqrt{n}\left\vert
n-1\right\rangle \left\langle n\right\vert =\hat{Z}_{\lambda}~. \label{zetaq1}%
\end{align}
Similarly, we have $\hat{\bar{\zeta}}=\hat{Z}_{\lambda}^{\dag}$. Let us now
quantize the classical observable $|\zeta|^{2}$ that, as it will be discussed
in the next section, represents a $\lambda$-deformation of the classical
relative angular momentum. One obtains:
\begin{align}
|\zeta|^{2}\mapsto &  \int_{\mathbb{C}}\frac{d^{2}\zeta}{\pi^{2}}%
\,\varpi_{\lambda}\left(  \left\vert \zeta\right\vert ^{2}\right)
\,\mathcal{E}_{\lambda}\left(  \left\vert \zeta\right\vert ^{2}\right)
\,|\zeta|^{2}~\left\vert \zeta\right\rangle \left\langle \zeta\right\vert
\nonumber\\
=  &  \sum_{n}(n+1)\exp\left[  \lambda(n+1)\right]  \left\vert n\right\rangle
\left\langle n\right\vert =\hat{Z}_{\lambda}\hat{Z}_{\lambda}^{\dag}%
=\partial_{\lambda}e^{\lambda(\hat{N}+1)}~. \label{zetaq3}%
\end{align}

Therefore, restoring the units,
\begin{equation}
\zeta\mapsto\hat{\zeta}=\sqrt{\frac{2\hbar}{\tilde{m}\tilde{\omega}}}%
\exp\left[  \frac{\lambda}{2}\left(  \hat{a}^{+}\hat{a}+1\right)  \right]
\hat{a}\ ,\;\bar{\zeta}\mapsto\hat{\zeta}^{+}=\sqrt{\frac{2\hbar}{\tilde
{m}\tilde{\omega}}}\hat{a}^{+}\exp\left[  \frac{\lambda}{2}\left(  \hat{a}%
^{+}\hat{a}+1\right)  \right]  ~, \label{zetaq2}%
\end{equation}
with the operator $\hat{a}$ and $\hat{a}^{+}$\ given in (\ref{r6}). Repeating
the procedure of the previous section, for the standard CS $\left\vert
z_{0}\right\rangle $, we obtain,%
\begin{equation}
z_{0}\mapsto r_{0-}=\sqrt{\frac{2\hbar}{\tilde{m}\tilde{\omega}}}\hat
{b}\ ,\;z_{0}^{\ast}\mapsto r_{0+}=\sqrt{\frac{2\hbar}{\tilde{m}\tilde{\omega
}}}\hat{b}^{+}~, \label{centerq1}%
\end{equation}
with the operator $\hat{b}$ and $\hat{b}^{+}$\ given in (\ref{r9}). Using the
relations (\ref{c6}) and the transformation (\ref{3}) we see that this CS
quantization reproduces the non-commutative relation (\ref{1}).

\subsubsection{Numerical analysis}

Following \cite{KowReP96} a criterion to test the closeness of the introduced
$\lambda$-CS (\ref{zetacs}) to the classical phase space, is verifying how
expectation value of the angular momentum operator approaches the respective
classical quantity. It can be done by the evaluation of the relative error
$e$,%
\begin{equation}
e\left(  \lambda,l\right)  =\frac{|\langle\hat{J}\rangle_{\zeta}-l|}{l}~,
\label{error}%
\end{equation}
with the expectation value of the angular momentum, in the units $\hbar
=\tilde{m}\tilde{\omega}/2=1$, given by%
\[
\langle\hat{J}\rangle_{\zeta}=\left\langle \zeta\right\vert \hat{J}\left\vert
\zeta\right\rangle =\frac{1}{\mathcal{E}_{\lambda}(|\zeta|^{2})}\sum
_{n=0}^{+\infty}\left\vert \zeta\right\vert ^{2n}\frac{\left(  2n+1\right)
}{n!}e^{-\frac{\lambda}{2}n\left(  n+1\right)  }\,.
\]
The parameter $\zeta$ is related to the classical angular momentum
$l=\tilde{m}\tilde{\omega}r^{2}=2r^{2}$ (where $r$ is the classical
radius)\ by%
\[
|\zeta|^{2}=\frac{l}{2}\exp\left(  \lambda\frac{l}{2}\right)  ~.
\]
As observed in \cite{KowRe05} the error computation (\ref{error}) by using
Kowalski-Rembielevski CS shows that the approximate equality $\langle\hat
{J}\rangle_{\zeta}\simeq l$ does not hold for arbitrary small $l$, being good
only for $l>1$. From Fig. 1 we see that, for our $\lambda$-CS, this
approximation can be improved, for $l\leq1$, by making $\lambda$ increase.%

{\parbox[b]{1.834in}{\begin{center}
\includegraphics[
height=1.8464in,
width=1.834in
]%
{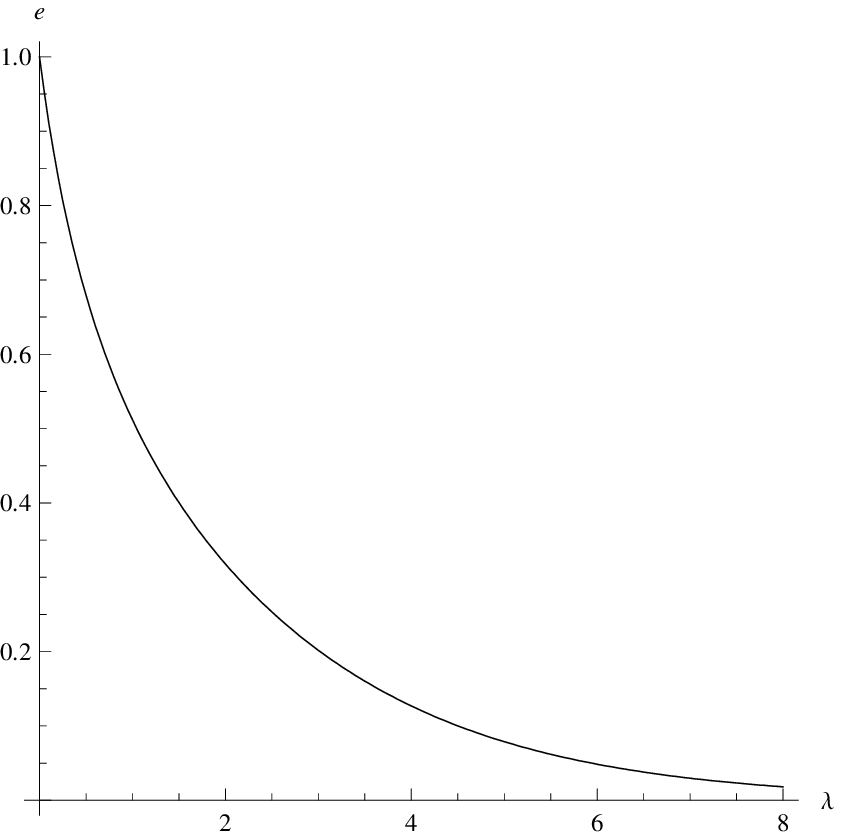}%
\\
\textsc{Fig.1}{\small -Error function }$e~${\small  as a function of }%
$\lambda~${\small  for }$\left\vert \zeta\right\vert =1${\small .}%
\end{center}}}%
\qquad%
\raisebox{-0.3379in}{\parbox[b]{1.8738in}{\begin{center}
\includegraphics[
height=1.8788in,
width=1.8738in
]%
{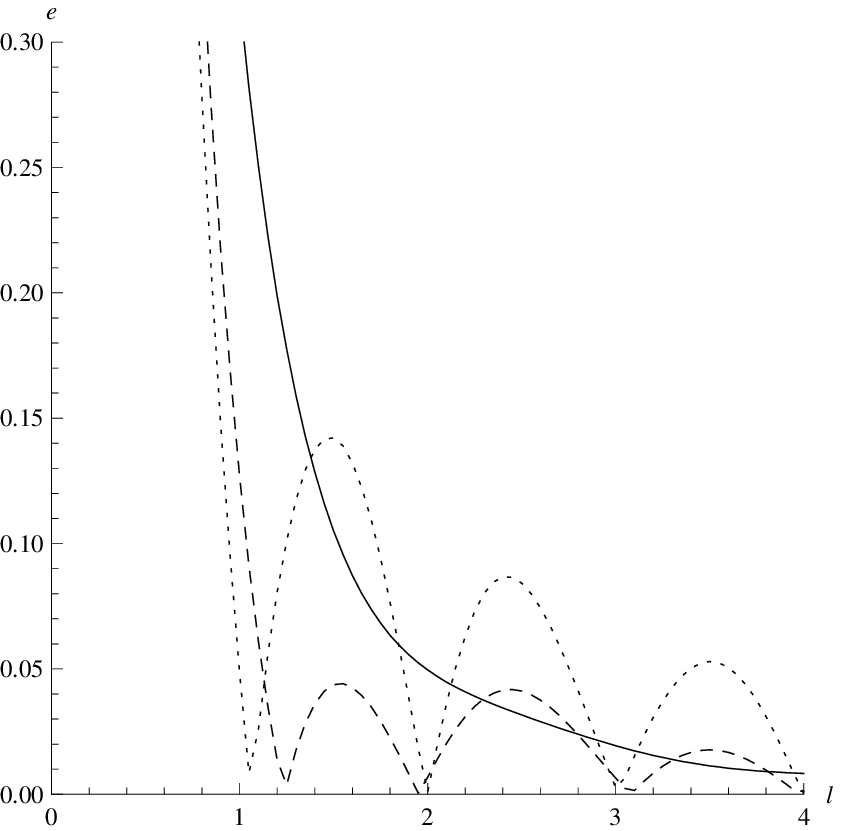}%
\\
\textsc{Fig.2}{\small -Error function }$e~${\small  as a function of }%
$l~${\small  for }$\lambda=2$ (solid line), $\lambda=4$ (dashed line) and
$\lambda=6$ (dotted line){\small .}%
\end{center}}}%

\bigskip

But this behavior is not shared by arbitrary values of $l>1$, as can be seen
in Figure 2: the error starts to oscillate as $l$ increases and, occasionally,
we can have $e\left(  \lambda,l\right)  >e\left(  \lambda^{\prime},l\right)  $
with $\lambda>\lambda^{\prime}$ for $l>1$. The approximation is better near
integer values assumed by $l$. This can be related to the fact that, in the
construction of the CS for a particle on a circle, as mentioned in
\cite{KowRe05}, the angular momentum can assume only integer values in the
boson case.

\subsubsection{Harmonic oscillator phase space}

In view of the commutation rule (\ref{comrulZ}) that illustrates a sort of
\textquotedblleft$\lambda$\textquotedblright\ deformation of the harmonic
oscillator, it is natural to consider the quantized version (\ref{zetaq3}) of
the classical observable $|\zeta|^{2}$ as a Hamiltonian $\hat{H}%
=\widehat{|\zeta|^{2}}=(\hat{N}+1)\exp(\lambda(\hat{N}+1))$ ruling the time
evolution of quantum states. We thus investigate the time evolution of the
quantized version $\hat{Z}_{\lambda}$, as found in (\ref{zetaq1}), of the
classical phase space point $\zeta\equiv(q+ip)/\sqrt{2}$, comparing it with
the phase space circular classical trajectories. This time evolution is well
caught through its mean value in coherent states $|\zeta\rangle$ (lower
symbol) \cite{GazBaG09}:
\begin{align}
&  \check{\zeta}\left(  t\right)  \overset{\mathrm{def}}{=}\left\langle
\zeta\right\vert e^{-i\hat{H}t}\hat{\zeta}e^{i\hat{H}t}\left\vert
\zeta\right\rangle \nonumber\\
&  =\frac{\zeta}{\mathcal{E}_{\lambda}(|\zeta|^{2})}\sum_{n=0}^{+\infty}%
\frac{\left\vert \zeta\right\vert ^{2n}}{x_{n}!}\,\exp\left[  {-i}\left(
{x_{n+2}-x_{n+1}}\right)  {t}\right]  \,, \label{timeev}%
\end{align}
with $x_{n}=ne^{\lambda}$. Setting the initial state $\zeta=1$, we plot the
phase-space ($\operatorname{Re}\zeta\times\operatorname{Im}\zeta$)\ for
different values of $\lambda$. For $\lambda=0$ we obtain a circle, as is
expected for the standard coherent states. For $\lambda\neq0$ the trajectories
are confined between two circles. The general behavior can be see in Figure 3,
where we set $\lambda=2$ (the circular CS of \cite{KowRe05}).%

{\parbox[b]{1.6986in}{\begin{center}
\includegraphics[
height=1.7036in,
width=1.6986in
]%
{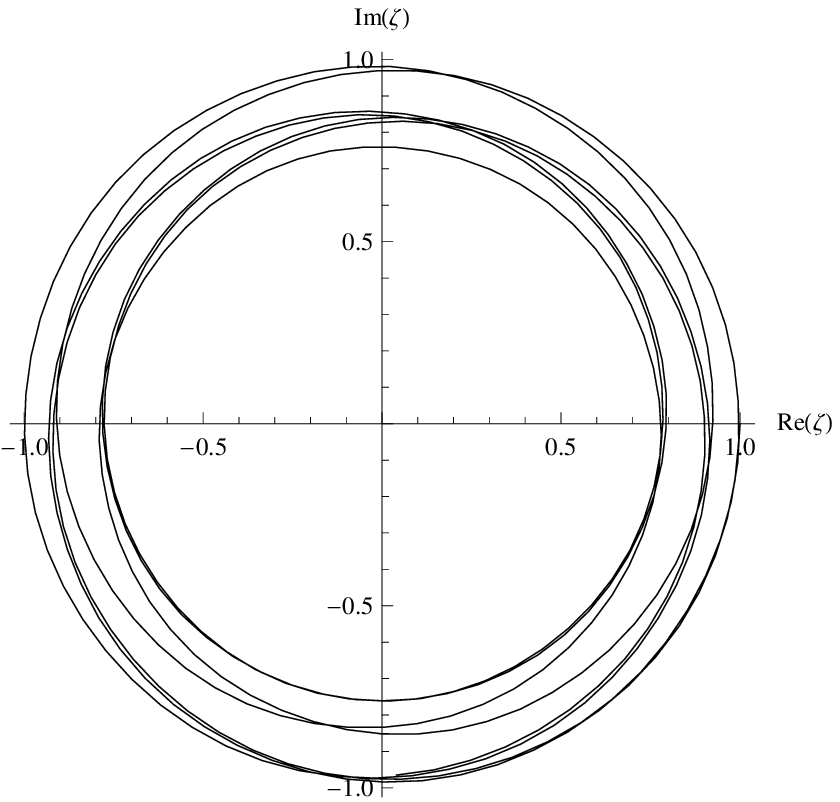}%
\\
\textsc{Fig.3}{\small -Phase trajectory for }$\lambda=2${\small , }$\zeta
=1~${\small  and }$0\leq t\leq8\pi${\small .}%
\end{center}}}%
\qquad%
{\parbox[b]{1.6953in}{\begin{center}
\includegraphics[
height=1.6903in,
width=1.6953in
]%
{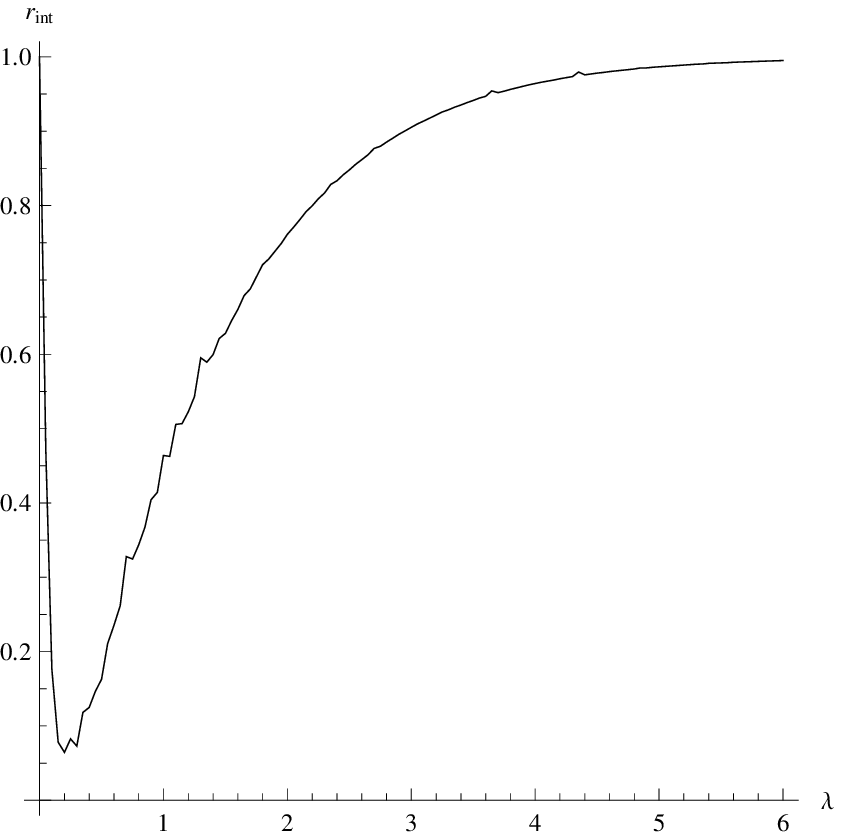}%
\\
\textsc{Fig.4}{\small -Dependence of the interno radius }$r_{\mathrm{int}}%
~${\small  with }$\lambda${\small .}%
\end{center}}}%

\bigskip

The dependence of the internal radius $r_{\mathrm{int}}$\ with the value of
$\lambda$ can be viewed in Figure 4. The internal radius decreases from $1$
(at $\lambda=0$) up to almost $0.05$ (for $\lambda\sim0.3$), which represents
the most squeezed version of our $\lambda$-CS. After that, it starts to
increase again. For $\lambda>6$ the trajectories become circular again, with a
period proportional to $e^{-\lambda}$.

\section{Conclusion}

We have studied a $\theta$-modified classical action for a charged particle in
a magnetic field. The canonical quantization of this model yields a quantum
mechanics for this non-commutative space. The classical theory and the quantum
theory are not gauge-invariant and obey some peculiarities for critical values
of the non-commutative parameter $\theta$. These values depend on the gauge.
In the symmetric gauge, the energy of the quantum particle depends on the
relation between the respective signs of $B$ and $\theta$. Thus, if we assume
that these quantities are independent, the difference between the energy
levels will change if we invert the direction of the magnetic field. We have
constructed the standard CS of particles in the $\theta$-modified quantum
theory and have shown that the mean values of the position operator coincide
with the \textquotedblleft classical" trajectories of the $\theta$-modified
classical theory. In addition, we have constructed a family of non-standard
circular CS parameterized by $\lambda\geq0$ and have used such states to
perform the Berezin-Klauder-Toeplitz quantization. As a result, we have
reproduced the $\theta$-modified quantum theory in the symmetric gauge. With
the help of numerical explorations, we have shown to what extent the mean
values of some physical quantities depend on the choice of the parameter
$\lambda$.

\begin{acknowledgement}
The authors are indebted to S. T. Ali (Univ. Concordia, Montreal) and F.
Bagarello (Univ. Palermo) for helpful comments and suggestions. M.C.B. thanks
FAPESP; D.M.G. thanks FAPESP and CNPq for permanent support. This work was
financed by CAPES-COFECUB, N%
${{}^\circ}$
PH 566/07.
\end{acknowledgement}


\begin{thebibliography}{99}                                                                                               %


\bibitem {Sch26}E. Schr\"{o}dinger, \emph{Der stetige \"{U}bergang von der
Mikro- zur Makromechanik}, Naturwiss. \textbf{14}, 664 (1926)

\bibitem {Gla63}R.J. Glauber, \emph{Photon Correlations}, Phys. Rev. Lett.
\textbf{10}, 84 (1963)

\bibitem {Sud63}E.C.G. Sudarshan, \emph{Equivalence of semiclassical and
quantum mechanical descriptions of statistical light beams}, Phys. Rev. Lett.
\textbf{10}, 277 (1963)

\bibitem {Klau60}J.R. Klauder, \emph{The Action Option and the Feynman
Quantization of Spinor Fields in Terms of Ordinary c-Numbers}, Ann. Phys.
\textbf{11}, 123 (1960)

\bibitem {Klau63}J.R. Klauder, \emph{Continuous-Representation Theory I.
Postulates of continuous-representation theory}, J. Math. Phys. \textbf{4},
1055 (1963)

\bibitem {ZhaFeG90}W.M. Zhang, D.H. Feng, and R. Gilmore, \emph{Coherent
states: Theory and some applications}, Rev. Mod. Phys. \textbf{62}, 867 (1990)

\bibitem {KlaSk85}J.R. Klauder, B.S. Skagerstam, \emph{Coherent States,
Applications in Physics and Mathematical Physics}, (World Scientific,
Singapore, 1985)

\bibitem {AliAnG00}S.T. Ali, J.P. Antoine, and J.-P. Gazeau, \emph{Coherent
states, wavelets and their generalizations. Graduate Texts in Contemporary
Physics}, (Springer-Verlag, New York, 2000)

\bibitem {Per72}A.M. Perelomov, \emph{Coherent states for arbitrary Lie
group}, Comm. Math. Phys. \textbf{26}, 222 (1972)

\bibitem {Ber2}F.~A. Berezin, \emph{The Method of Second Quantization} (Nauka,
Moscow, 1965); \emph{Introduction to Algebra and Analysis with Anticommuting
Variables} (Moscow State University Press, Moscow, 1983); \emph{Introduction
to Superanalysis} (D. Reidel, Dordrecht, 1987)

\bibitem {MalMa68}I.A. Malkin and V.I. Man'ko, \emph{Coherent States of a
Charged Particle in a Magnetic Field}, Zh. Eksp. Teor. Fiz. \textbf{55}, 1014
(1968) [Sov. Phys. - JETP \textbf{28}, no.3, 527 (1969)]

\bibitem {LoyMoS89}G. Loyola, M. Moshinsky, and A. Szczepaniak, \emph{Coherent
states and accidental degeneracy for a charged particle in a magnetic filed},
Am. J. Phys. \textbf{57}, 811 (1989)

\bibitem {SchMo03}D. Schuch and M. Moshinsky, \emph{Coherent states and
disisipation for a motion of a charged particle in a magnetic field}, J. Phys.
A \textbf{36}, 6571 (2003)

\bibitem {KowRe05}K. Kowalski and J. Rembielinski, \emph{Coherent state of a
charged particle in a uniform magnetic filed}, J. Phys. A \textbf{38}, 8247 (2005)

\bibitem {KowReP96}K. Kowalski, J. Rembielinsk, and L.C. Papaloucas,
\emph{Coherent states for a quantum particle on a circle}, J. Phys. A
\textbf{29}, 4149 (1996)

\bibitem {GazBaG09}J.-P. Gazeau, M.C. Baldiotti, and D.M. Gitman,
\emph{Coherent states of a particle in a magnetic field and the Stieltjes
moment problem}, Phys. Lett. A \textbf{373}, 1916 (2009)

\bibitem {Hor02}P.A. Horv\'{a}thy, \emph{The non-commutative Landau problem},
Ann. Phys. \textbf{299}, 128 (2002)

\bibitem {AlGoKP08}P.D. Alvarez, J. Gomis, K. Kamimura, and M.S. Plyushchay,
\emph{Anisotropic harmonic oscillator, non-commutative Landau problem and
exotic Newton--Hooke symmetry}, Phys. Lett. B \textbf{659}, 906 (2008)

\bibitem {GelGaS09}J.B. Geloun, S. Gangopadhyay, F.G. Scholtz, \emph{Harmonic
oscillator in a background magnetic field in noncommutative quantum
phase-space}, arXiv:0901.3412 (2009)

\bibitem {DelDUGL07}F. Delduc, Q. Duret, F. Gieres, and M. Lefran\c{c}ois,
\emph{Magnetic fields in noncommutative quantum mechanics}, arXiv:0710:2239v1 (2007)

\bibitem {SchGoHR08}F.G. Scholtz, L. Gouba, A. Hafver and C.M. Rohwer,
\emph{Formulation, interpretation and application of non-commutative quantum
mechanics}, J. Phys. A \textbf{42}, 175303 (2009)

\bibitem {Chaichian1}M. Chaichian, M.M. Sheikh-Jabbari, A. Tureanu,
\emph{Hydrogen atom spectrum and the Lamb shift in noncommutative QED}, Phys.
Rev. Lett. \textbf{86}, 2716 (2001)

\bibitem {AdoBaCGT09}T.C. Adorno, M.C. Baldiotti, M. Chaichian, D.M. Gitman
and A. Tureanu, \emph{Dirac Equation in Noncommutative Space for Hydrogen
Atom}, arXiv:0904.2836 (2009)

\bibitem {GitKu08}D.M. Gitman and V.G. Kupriyanov, \emph{Path integral
representations in noncommutative quantum mechanics and noncommutative version
of Berezin-Marinov action}, Eur. Phys. J. C \textbf{54}, 325 (2008)

\bibitem {Deriglazov}A.A. Deriglazov, \emph{Noncommutative version of an
arbitrary nondegenerated mechanics}, hep-th/0208072 (2002)

\bibitem {Duval}C. Duval and P.A. Horv\'{a}thy, \emph{Exotic Galilean symmetry
in the non-commutative plane and the Hall effect}, J. Phys. A \textbf{34},
10097 (2001)

\bibitem {GitTy90}D.M. Gitman and I.V. Tyutin, \emph{Quantization of Fields
with Constraints} (Springer-Verlag 1990)

\bibitem {IFUSP1648}D.M. Gitman and V.G. Kupriyanov, \emph{Gauge invariance
and Lagrangian formulation for particle theory on noncommutative space},
University of S\~{a}o Paulo, Report number \textbf{IFUSP 1648/2009} (2009)

\bibitem {JohLi49}M.H. Johnson and B.A. Lippmann, \emph{Motion in a Constant
Magnetic Field}, Phys. Rev. \textbf{76}, 828 (1949)

\bibitem {Ber71}F.A. Berezin, \emph{Wick and anti-Wick operator symbols}, Mat.
Sb. \textbf{86}, 578 (1971) [English transl. in Math. USSR-Sb. \textbf{15},
577 (1971)]
\end{thebibliography}
\end{document}